\documentclass[3p,times]{elsarticle}

\usepackage{ecrc}

\volume{00}

\firstpage{1}

\journalname{Nuclear Physics A}

\runauth{Xin-Nian Wang}

\jid{npa}

\jnltitlelogo{Nuclear Physics A}

\usepackage{amssymb,amsmath}
\biboptions{square,comma,numbers,sort&compress}

\usepackage[figuresright]{rotating}

\newcommand{\ba}{\begin{eqnarray}}
\newcommand{\ea}{\end{eqnarray}}

\begin{document}

\begin{frontmatter}



\dochead{}
%% e.g. \dochead{17th International Conference on Dynamical Processes in Excited States of Solids}

\title{What hard probes tell us about the quark-gluon plasma: Theory}


\author[a1,a2]{Xin-Nian Wang}

\address[a1]{Institute of Particle Physics and Key Laboratory of Quark and Lepton Physics (MOE),
Central China Normal University, Wuhan 430079, China}
\address[a2]{Nuclear Science Division Mailstop 70R0319, Lawrence Berkeley National Laboratory, Berkeley, California 94740, USA}

\begin{abstract}
 In the study of the quark-gluon plasma in high-energy heavy-ion collisions, hard and electromagnetic (EM) processes play an essential role as probes of the properties  of the dense medium. They can be used to study a wide range of properties of the dense medium in high-energy heavy-ion collisions, from space-time profiles of the dense matter, bulk transport coefficients to EM responses and the jet transport parameter. I review in this talk these medium properties, how they can be studied through hard and EM probes and the status of recent theoretical and phenomenological investigations. 

\end{abstract}

\end{frontmatter}

% \linenumbers

\section{Introduction}
%\label{intro}
%\hspace{1cm}\\ \\

Many of us don't believe in metaphors. However, the distances that we all have traveled from around the world to come to this meeting near Cape Town of South Africa figuratively symbolize how far we have come in the pursuit of quark-gluon plasma (QGP) using hard probes. In the search for QGP in heavy-ion collisions, hard and electromagnetic probes have become increasingly important. Starting as some mere theoretical ideas in the early days of heavy-ion collisions at Brookhaven AGS and CERN SPS energies where these processes are quite rare, they become the bread-and-butter and powerful tools for the study of the properties of the strongly interacting QGP in high-energy heavy-ion collisions at Brookhaven RHIC and CERN LHC. At such high colliding energies, cross sections of hard processes are so large that these processes occur several and tens times in each central heavy-ion collision. The energy scales accessible become large enough that it becomes possible to carry out precision studies of the medium properties with hard probes relative to the baseline cross sections one can calculate in perturbative QCD (pQCD) \cite{hpc} or measure in p+p and p+A collisions.  These are the focus of discussions in the Hard Probes Conferences in the era of heavy-ion collisions at unprecedented high energies.

To answer the question in the title of this opening talk as requested by the organizers, it is important to take a look back at what properties of QGP that we are interested in and what can be studied in high-energy heavy-ion collisions. One in generally can divide these properties into the categories of bulk medium variables, bulk transport properties, electromagnetic response and medium response to hard and propagating partons. Because of the transient nature of the dense medium in heavy-ion collisions, one has to reply on signals from the internal interactions during the dynamic evolution of the dense matter to study its properties. The study often involves theoretical modeling of the dynamic evolution of and interaction inside the dense matter and comparisons to experimental data for extraction of its properties.

\section{Bulk medium and transport coefficients}

The first and most basic set of properties of the dense matter that we want to know in heavy-ion collisions are the space-time profiles of the bulk medium in terms of local temperature and fluid velocity. Our knowledge of these properties forms the foundation that we have to reply on for the investigation of all other medium properties. Since the bulk medium and its collective expansion involve mostly soft particles, we consider them as soft probes of the dense medium. The most studied effective theory for the space-time evolution of the bulk medium is the relativistic hydrodynamical model. With the input of initial conditions of the spatial distribution of the energy density and fluid velocity profiles and the equation of state (EoS), one can solve the relativistic hydrodynamic equations for the evolution of the bulk medium. The final hadron spectra on a hype-surface at a given freeze-out temperature will depend most dominantly on the initial time (thermalization time), initial energy density distribution, EoS and the freeze-out temperature. Comparisons to the experimental data on hadron spectra can provide some constraints on initial conditions and EoS but not sufficient to determine the EoS  \cite{Shen:2010uy}. The most common practice is to take the EoS as an input from lattice QCD simulations in terms of a parametrized form \cite{Huovinen:2009yb}. The experimental data on final hadron spectra and their azimuthal anisotropies will then provide sensitive constraints on the initial conditions and the transport properties of the medium, in particular the shear viscosity,
\begin{equation}
\eta =\lim_{\omega\rightarrow 0}\frac{1}{2\omega}\int dt dx e^{i\omega t}\langle \left[T_{xy}(0),T_{xy}(x)\right]\rangle.
\end{equation}
With reasonable models for the initial conditions, the hydrodynamic simulations of heavy-ion collisions can describe the observed anisotropic flows at both RHIC and LHC with the value of shear viscosity to entropy density ratio $\eta/s\approx 0.08-0.2$ \cite{Song:2010mg,Gale:2012rq}. This is one of the few transport properties of the bulk medium one can extract from comparisons of experimental data and hydrodynamical model simulations. For the description of other hard probes in heavy-ion collisions in a hydrodynamic background, effects due to viscous corrections with these values of transport coefficients are small. For this purpose, 3+1D ideal hydrodynamical model will also be sufficient. Within the framework of the JET Collaboration, a numerical code called iEBE has been developed and made available for public use \cite{iebe}. It can generate space-time profiles of hydrodynamic evolution of bulk medium with fluctuating initial conditions in high-energy heavy-ion collisions which can be used for calculations of EM emission and propagation of hard probes.

\section{EM emission, quarkonium suppression and color screening}

Interaction of quarks in a dynamic system in heavy-ion collisions can lead to emission of photons and dileptons which can escape the system with approximately no further interaction, therefore carrying important information of the medium and the interaction inside. This is vey similar to using deeply inelastic scattering (DIS) and Drell-Yan processes for the study of parton distributions inside a nucleon or nucleus. What these EM emissions probe is the EM response function of the medium,
\begin{equation}
W_{\mu\nu}(q)=\int \frac{d^4x}{4\pi} e^{iq\cdot x}\langle j_\mu(0) j_\nu(x)\rangle,
\end{equation}
which is essentially the thermal assemble average of the current correlation function and determines the local EM emission rate for a photon (or a virtual photon) with four-momentum $q$ \cite{McLerran:1984ay}.  One can calculate this EM emission rate within pQCD at finite temperature. At the leading order (LO) of the strong coupling constant $\alpha_{\rm s}$, one already has to resum soft gluon emissions whose momenta are parametrically smaller than the temperature $gT\ll T$ leading to the so-called Hard Thermal Loop resummation that regulates the logarithmic divergence in the contribution to the rate from soft gluon emissions. Collinear gluon emissions induced by multiple small angle scattering including LPM interference effect can also be resumed to contribute at LO to the EM emission rate. Recent calculations \cite{Ghiglieri:2014vua} have included contributions at the next-leading order (NLO) ($g^3$) from loop corrections and contributions from other mistreated regions of phase-space in the LO calculations. Numerically, the net NLO corrections increase the total rate about 20\% over the LO result. The EM emission rate for virtual photons (or dileptons) has also been calculated in lattice QCD \cite{ding} in quenched configurations. The rate can be matched to HTL resummed LO results at large photon energy but shows a few orders of magnitude enhancement at low momentum, illustrating the importance of non-perturbative contributions. The enhancement is also approximately independent of the temperature above the QGP phase transition. It will be interesting to see whether and how inclusion of dynamic quarks in the lattice QCD calculation will change the result.

The most significant experimental result from RHIC on EM emission is the observed large enhancement of soft photons and the large azimuthal anisotropy $v_2$ \cite{Adare:2011zr}. The enhanced low transverse momentum photons indicate a high temperature  and radial flow at the time of emission and large $v_2$ implies a late emission time when the interacting matter has developed significant anisotropy in the phase-space due to collective expansion. Theoretical calculations with the HTL pQCD rate in QGP found that dominant contributions to direct photon spectra at low $p_T\le 2$ GeV/$c$ come from the hadronic phase therefore has large $v_2$ \cite{vanHees:2011vb,Heinz:2014uga}.  But it is still smaller than the experimental data. Inclusions of viscous corrections to the photon emission rate has been recently considered \cite{Shen:2014cga}. They however further reduce the final azimuthal anisotropy of the photon spectra. Such tension between experimental data and pQCD results might imply the importance of non-perturbative contributions to the EM emission rate which still awaits for full lattice QCD calculations.

\begin{figure}[t]
\centerline{\includegraphics[width=7.8cm]{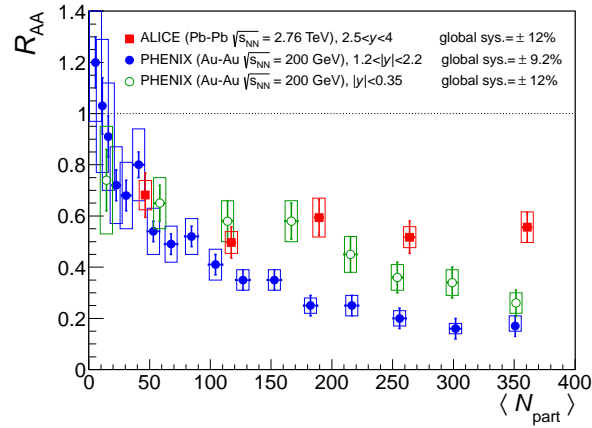}}
 \caption{ (color online) Comparison of the suppression factors for J/$\psi$ production in Pb+Pb collisions at LHC and Au+Au collisions at RHIC  \cite{Abelev:2012rv} as a function of the number of participant nucleons.
 \label{fig:jpsi}}
\end{figure}

Study of the EM response of the medium to a virtual photon can be extended to the region of hadron resonances which can provide information on medium modification of hadron states due to chiral symmetry restoration in the light quark sector \cite{Heinz:2014uga}. In the region of heavy-quark resonances,  suppression of quarkonia in QGP due to color screening of the confining potential between heavy quark-anti-quark pairs has been suggested as a good probe of the color deconfinement in QGP \cite{Matsui:1986dk}. Development of techniques such as the Maximum-Entropy Method has allowed the study of the spectra function of heavy vector mesons in lattice QCD \cite{Asakawa:2000tr}. Results of the latest lattice QCD study suggests that both S-wave and P-wave states of charmonium melt at $T\ge 1.5 T_c$ \cite{ding,moscy}. While the P-wave states of bottonium melt immediately at $T \approx T_c$ the S-wave states remain up to $2T_c$. Charmonium suppression beyond the cold nuclear effects in the initial production has been observed in heavy-ion collisions at SPS, RHIC and LHC. Experimental data in the colliding energy, centrality and $p_T$ dependence of the final charmonium spectra indicate a strong suppression at SPS energy and above. However, starting at the RHIC energy such suppression due to color screening is partially compensated by regeneration of charmonia from recombination of increasing number of open charm quarks  and anti-quarks from initial production. This leads to less total suppression of charmonium yield at LHC than at RHIC as shown in Fig.~\ref{fig:jpsi}.  A consistent model calculation of the final quarkonium yield should include both the dissociation and regeneration of quarkonium that depends on the color-screened heavy quark potential \cite{Zhao:2011cv} which can be extracted from lattice QCD calculations \cite{ding,moscy}. Because of the interplay between suppression and regeneration, there will also be a unique centrality and energy dependence of the average transverse momentum squared of the final quarkonium \cite{Zhou:2013aea}.  Systematic studies of these features can then provide constraints on the color screening in QGP.

%%%%%%%%%%%%%%%%%%%%%%%%% Fig 1 %%%%%%%%%%%%%%%%%%%%%%%%%%%%%%%%%

\section{Jet transport parameter}

In addition to soft and EM probes of the medium, one can also use initially produced energetic partons as hard probes of the QGP as they propagate through the dense medium in heavy-ion collisions. These energetic partons will suffer multiple scattering in the medium which will lead to transverse momentum broadening and parton energy loss. The interaction between a propagating parton and the medium can be characterized by the transverse momentum broadening squared per unit length which in turn can be related to the ensemble average of the Lorentz force exerted on the parton by the chromo field of the medium,
\begin{equation}
\hat q_R= \int d^2q_T q_T^2 \sum_a\rho_a\frac{d\sigma_{Ra}}{d^2q_T} =\frac{4\pi^2\alpha_s C_R}{N_c^2-1}\int\frac{dy^-}{\pi}\langle F^{\sigma +}(0) F_{\sigma}^{\,+}(y)\rangle.
\end{equation}
This quantity is often referred to as the jet transport parameter for a propagating parton in color representation $R$ and is one of the fundamental properties of the QGP medium that we would like to extract from experimental data. This quantity has been calculated within the framework of HTL pQCD at NLO \cite{CaronHuot:2008ni}. However, the uncertainty in the choice of the kinetic cut-off for the transverse momentum integration is still very large. There are attempts to evaluate this quantity with lattice QCD calculations using either the analytic continuation of the imaginary-time formulation \cite{Majumder:2012sh} or dimensionally reduced effective theory (electrostatic QCD) \cite{Panero:2013pla}. The EQCD method computes the contribution from soft modes and matches the result to the NLO pQCD result at higher transverse momentum. Therefore, the final total contribution is also subject to the uncertain in the choice of the cut-off. 

Multiple scatterings in medium lead to both elastic and radiative energy loss for a propagating parton which in turn can result in jet quenching or suppression of large $p_T$ hadron and jet spectra in heavy-ion collisions \cite{Wang:1991xy}. This predicted phenomenon has been observed in high-energy heavy-ion collisions at both RHIC and LHC. One not only has observed the suppression of single inclusive hadron spectra at large $p_T$ by a factor of 5 but also the suppression of back-to-back dihadron and gamma-hadron correlations relative to the spectra and correlation in p+p and p+A collisions. These patterns of jet quenching are consistent with the picture of parton energy loss and point to strong interaction between energetic jet shower partons and the medium.   Parton energy loss and the suppression of hadron and jet spectra, which are dominated by induced gluon emission, are directly controlled by the jet transport parameter along the trajectory of parton propagation. Through comparisons of experimental data and theoretical calculations of hadron and jet suppression one can extract the jet transport parameter and study its temperature and energy dependence. 

For a precision extraction of the jet transport parameter from experimental data on jet quenching, one needs to combine jet transport simulations with realistic 3+1D hydrodynamic modeling of the bulk medium evolution and final hadronization of jet shower partons. This is the purpose of many coordinated efforts such as the JET Collaboration. Within a coordinated effort, a general framework is developed for the numerical implementation of different approaches and improvements of jet transport simulations for the assessment of theoretical uncertainties. One has to combine the fragmentation and parton recombination mechanisms in a general description of jet hadronization in a medium. Initial conditions and bulk transport coefficients for hydrodynamical models should be constrained by the bulk hadron spectra and their azimuthal anisotropies. 

\begin{figure}[t]
\centerline{\includegraphics[width=7.8cm]{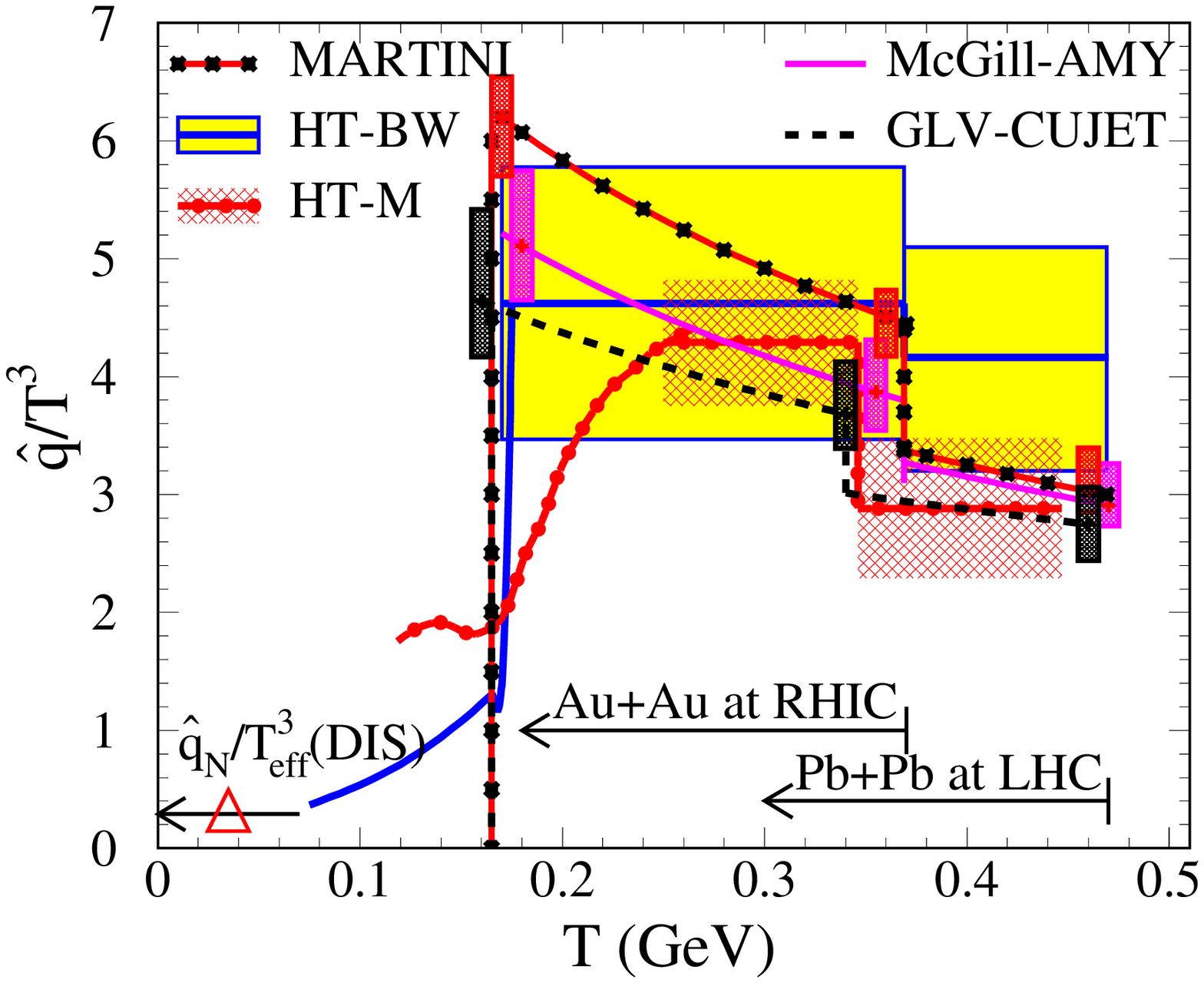}
\includegraphics[width=7.8cm]{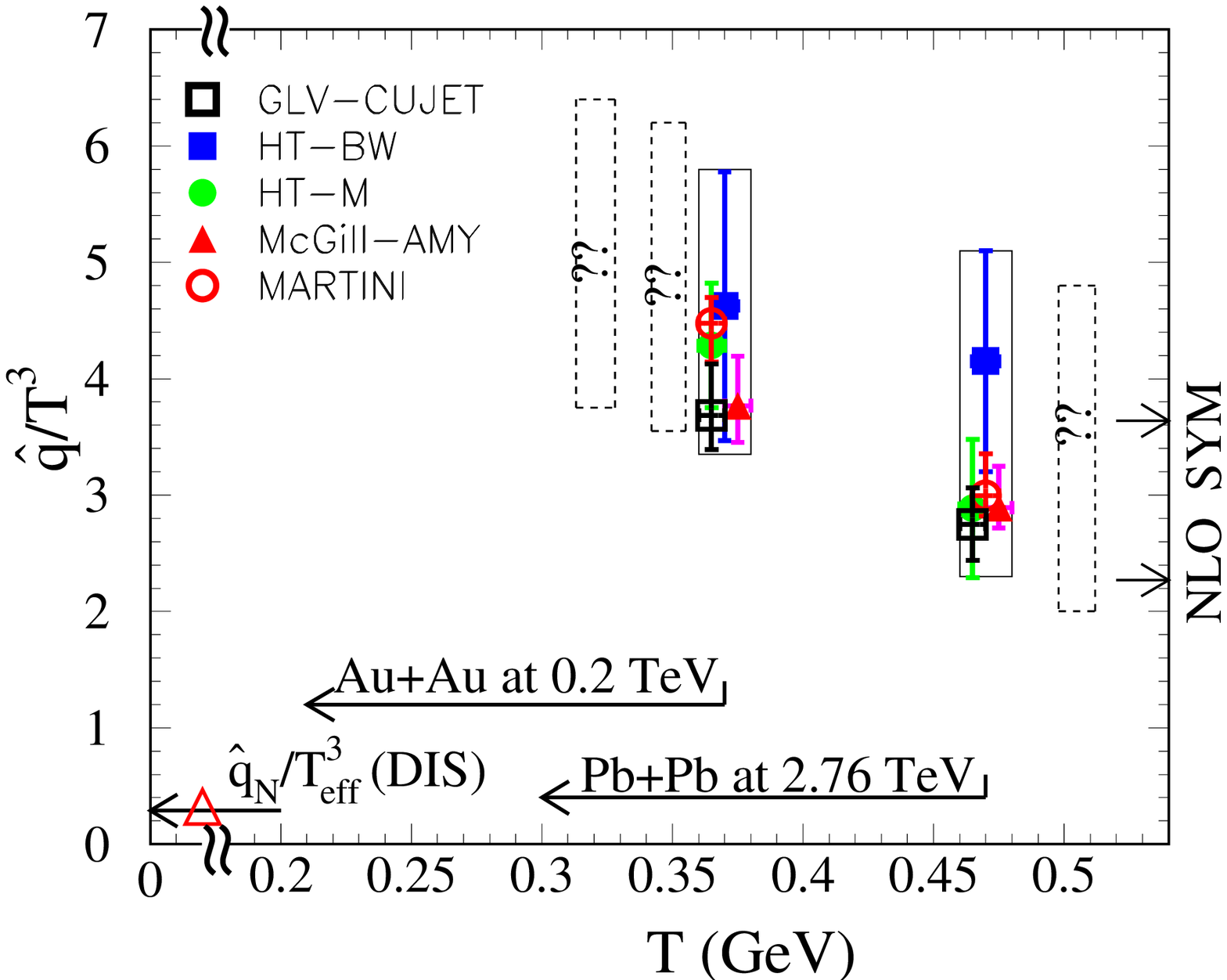}}
 \caption{ (color online) (left panel) The scaled jet transport parameter $\hat q/T^3$  for an initial quark jet with energy $E=10$ GeV as a function of temperature from fits to the suppression factors of single hadron spectra in central Au+Au collisions at RHIC and Pb+Pb collisions at LHC  \cite{Burke:2013yra} within five different approaches to parton energy loss. Boxes indicate errors from model fits. (right panel) Values of scaled jet transport parameters $\hat q/T^3$ from model fits to single hadron suppression, with dashed boxes indicating expected values from future phenomenological analyses. Two arrows on the right axis indicate NLO SYM results. The arrow at the left bottom corner indicates the value in cold nuclei from DIS experiments.
 \label{fig:qhat}}
\end{figure}

As the first result from the JET Collaboration \cite{Burke:2013yra}, a JET package is developed to combine five different semi-analytic formulations of medium modification of single hadron spectra with bulk medium evolution from hydrodynamic simulations. The five semi-analytic formulations in the first comparative study include GLV-CUJET, HT-M, HT-BW, MARTINI and McGill-AMY.  Through $\chi^2$ fits to the experimental data on the medium modification factors $R_{AA}(p_T)$ for single hadron spectra in Au+Au collisions at RHIC ($\sqrt{s}=200$ GeV) and Pb+Pb collisions at LHC ($\sqrt{s}=2.76$ TeV), values of the jet transport parameter $\hat q_0$  at the center of the most central collisions at an initial time $\tau_0=0.6$ fm/$c$ are extracted or calculated from each model.   From these values of extracted jet transport parameters, as shown in the left panel of Fig.~\ref{fig:qhat}, one can determine the range of its values $\hat q_0/T^3\approx 4.6\pm 1.2 $ and  $3.7\pm1.4$ at the highest temperatures in the most central Au+Au collisions at RHIC and Pb+Pb collisions at LHC, respectively, for a quark with an initial energy of 10 GeV.  The corresponding absolute values are $\hat q_0\approx 1.2\pm0.3$ and $1.9\pm0.7$ at $T=370 $ MeV and 470 MeV, respectively. These values are many orders of magnitude higher than in a cold nucleus as extracted from jet quenching in DIS off large nuclei (indicated by the arrow on the left-bottom corner). They are consistent with LO pQCD calculations of the transverse momentum broadening of a propagating parton in a QGP medium. They are also comparable to the NLO supersymmetric Yang-Mills results (as indicated by the arrows on the right axis in the right panel of Fig.~\ref{fig:qhat}).  If one can carry out similar analyses of future LHC higher energy ($\sqrt{s}=5.5$ TeV) and RHIC beam scan energy ($\sqrt{s}=20-200$ GeV) data, one should be able to constrain the temperature dependence of the jet transport parameter $\hat q$, as shown in the right panel of Fig.~\ref{fig:qhat}. Similar systematic studies of hadron suppression in jets with fixed initial energy or gamma-hadron correlations should also enable one to extract the energy dependence of $\hat q$.

%%%%%%%%%%%%%%%%%%%%%%%%% Fig 1 %%%%%%%%%%%%%%%%%%%%%%%%%%%%%%%%%

\section{New theoretical developments}

For further improvements of theoretical studies of jet transport and propagation, one needs to consider multiple gluon emissions induced by jet-medium interaction. Resummation of these multiple gluon emissions in the leading logarithmic approximation is equivalent to solving medium modified DGLAP evolution equations which can also be carried out by Monte Carlo simulations together with elastic scatterings in medium. Many such Monte Carlo models have been developed in the last few years, such as JEWEL, Linearized Boltzmann Transport (LBT) model, MARTINI and Parton Cascade Model (PCM). Going beyond the leading logarithmic approximation, one needs to consider color interferences between multiple gluon emissions. Recent studies \cite{Iancu, Salgado} show, however, that color coherence is rapidly lost in medium due to color-exchanges with medium through soft interactions. Given a propagation length $L$, independent emissions are enhanced by $L/\tau_f$, $\tau_f=\sqrt{2\omega/\hat q}$, and become dominant in later evolution. 

During parton propagation, the energy loss of the leading parton is transferred to radiated gluons and recoiled thermal partons, which in turn will further interact and propagate in the medium and become jet-induced medium excitations. Such jet-induced medium excitations can provide important information about medium transport properties. Furthermore, they also influence the underlying background in a jet event and the transverse energy distribution inside a reconstructed jet.  A recent study within the LBT model shows that recoiled thermal medium partons within a jet cone ($R=0.3$) can reduce the effective final jet energy loss by 20\% \cite{Wang:2013cia}. It remains a challenge experimentally to incorporate the effect of these recoiled thermal partons and the jet-induced medium excitation in the jet reconstruction. Single hadron spectra, dihadron and gamma-hadron correlations may still be the most effective tools for the study of jet quenching and in particular jet-induced medium excitations.

Uncertainties in existing theoretical studies of parton energy loss and medium modification of hadron spectra lie in the collinear approximation of the induced gluon emission and therefore the large-angle cut-off dependence of the final results. Such uncertainties can be reduced only by inclusion of large angle gluon emissions within the complete NLO calculation of induced gluon radiation and initial production of the hard parton. The first such study has recently been completed in DIS off large nuclei and DY process in p+A collisions \cite{Kang:2013raa}. Factorization was verified at NLO and twist-4 and therefore the universality of the jet transport parameter in the medium. A QCD evolution equation is also derived for the jet transport parameter which will determine its scale dependence. Similar efforts have also been carried out to calculate NLO corrections to the transverse momentum broadening \cite{mehta-tani,wu}

\section{Summary}

In the last decade since the first observation of jet quenching at RHIC, hard and EM probes have provided powerful tools for the study of the properties of the strongly interacting QGP in high-energy heavy-ion collisions. Phenomenological analyses of experimental data on EM emission, suppression of quarkonium and jet quenching point to an interacting QGP in the center of heavy-ion collisions at RHIC and LHC with high initial temperature, screening of the confining potential between quark and anti-quarks, frequent recombination of heavy quarks and anti-quarks into regenerated quarkonia and a value of jet transport parameter that is consistent with the pQCD description of interaction between an energetic parton and thermal QGP medium.  Future precision studies rely on theoretical understandings of multiple gluon emissions, color-decoherence, jet-induced medium excitations and NLO corrections to parton energy loss and $p_T$ broadening. One of the final goals of the study is to map out both the temperature and energy dependence of the jet transport parameter which complement the bulk transport coefficients determined from soft probes.

This work is supported by the NSFC under Grant No. 11221504,  China MOST under Grant No. 2014DFG02050, the Major State Basic Research Development Program in China (No. 2014CB845404), U.S. DOE under Contract No. DE-AC02-05CH11231 and within the framework of the JET Collaboration.



%% References


\bibliographystyle{elsarticle-num}
\bibliography{<your-bib-database>}



\end{document}